\documentclass[runningheads]{svmult}

\usepackage{graphicx}  
\usepackage{subeqnar}  
\usepackage{multicol}  
\usepackage{physprbb}  

\usepackage{amsmath}   
\usepackage{url}       
\usepackage{amssymb}
\usepackage{subfigure}
\begin{document}
%
%
\begin{titlepage}
\vspace*{3truecm}
\begin{center}
\begin{sffamily}
{\Large What do phase space methods tell us about\\[0.3\baselineskip]
disordered quantum systems?}\\[2\baselineskip]
{\large Gert-Ludwig Ingold, Andr{\'e} Wobst, Christian Aulbach,\\
and Peter H{\"a}nggi}\\[0.3\baselineskip]
Institut f{\"u}r Physik, Universit{\"a}t Augsburg, D-86135
Augsburg\\[22\baselineskip]
\end{sffamily}
to be published in\\
``The Anderson Transition and its Ramifications --
Localisation, Quantum Interference, and Interactions'',\\
Lecture Notes in Physics, \url{http://link.springer.de/series/lnpp/}\\
\copyright\ Springer Verlag, Berlin-Heidelberg-New York
\end{center}
\end{titlepage}
\setcounter{page}{1}
%
%
%
\title*{What do phase space methods tell us about disordered quantum systems?}
\toctitle{What do phase space methods tell us about disordered quantum systems?}
%
%
\titlerunning{Phase space and disordered systems}
%
\author{%
Gert-Ludwig Ingold\and
Andr{\'e} Wobst\and
Christian Aulbach\and
Peter H{\"a}nggi}
\authorrunning{Gert-Ludwig Ingold et al.}
%
%
\institute{Institut f{\"u}r Physik, Universit{\"a}t Augsburg,
D-86135 Augsburg, Germany}

\maketitle              


\section{Introduction}
At a summer school held in fall 1991 at the PTB in Braunschweig,
Bernhard Kramer proposed a book project that should encompass novel solid
state research topics ranging from quantum transport to quantum
chaos. This very book finally appeared at the end of 1997 under
the title ``Quantum transport and dissipation''
\cite{ingold:qtad98}. Compiled between the book covers is a series
of topics, some of which are usually not connected with each other in
present day's research. The main subject in the first chapter is
coherent transport in disordered systems. The last chapter, on the
other hand, dwells on concepts such as  phase space, Wigner and
Husimi functions, and alike.  In any case, in the description of
disordered systems we do not find many works that explicitly make
use of phase space concepts. This is so, although the existence of
a mapping between the Anderson model and the kicked rotor
\cite{ingold:fishm82}  indeed suggests that methods employed in
quantum chaos may advantageously be used to elucidate the
physics at work in  disordered quantum systems.

There exists, however, another, almost obvious physical motivation
to utilize the powerful phase space concepts to study disordered
quantum systems: as a function of the disorder strength, the
nature of the eigenstates changes from ballistic to localized
behavior, containing possibly in-between a diffusive regime
\cite{ingold:krame93}. In
the case of ballistic transport it is appropriate to think in terms 
of plane waves which are occasionally scattered by the weak
disorder potential. Then a momentum space description clearly
imposes itself. For strong disorder, however, the natural
physical space is  real space. A unified description, being valid
at arbitrary disorder strength, can thus  be achieved by focusing
 on suitable (quantum) phase space concepts.

\section{Phase space methods in quantum mechanics}
When an attempt is made to represent a quantum state in phase
space, there are two observations which deserve our attention.
First, phase space combines conjugate quantum observables which
brings into play a Heisenberg uncertainty relation. While in classical 
mechanics, the state of a point particle can be described in phase space 
with infinite precision, this  no longer holds true in quantum mechanics. 
Therefore, we expect some difficulties if for quantum systems one attempts 
to obtain perfect resolution in phase space. Second, the full 
information about a quantum state is  contained already in the 
corresponding wave function which can either be expressed in position 
or momentum space representation. Thus, the phase space representation 
of a quantum state cannot provide more information than the wave function 
itself. This information, however, may be presented in a more 
advantageous form, as we shall demonstrate next.

\subsection{The Wigner function}
As a first possibility to define a phase space representation of a
quantum state possessing the wave function $\psi(x)$ we introduce
the Wigner function
\begin{equation}
W(x,k) = \int\D y\E^{\I
ky}\psi^*(x+\frac{y}{2})\psi(x-\frac{y}{2})\,.
\label{ingold:eq:wigner}
\end{equation}
Here, and in the following we use the wave number $k$ instead of
the momentum $p=\hbar k$. The definition (\ref{ingold:eq:wigner})
is useful and appropriate as can be seen by considering the moments of 
position and momentum observables. Upon evaluation of the momentum integral, 
one readily finds that
\begin{equation}
\int\frac{\D x\D k}{2\pi}x^nW(x,k) = \int\!\D x\,x^n\vert\psi(x)\vert^2.
\label{ingold:eq:xmoment}
\end{equation}
With a little more effort one further can establish that
\begin{equation}
\begin{aligned}
\int\frac{\D x\D k}{2\pi}k^nW(x,k) &=
\int\!\!\D x\int\!\!\D y\int\frac{\D k}{2\pi}\left[(-\I)^n\frac{\D^n}{\D y^n}
\E^{\I ky}\right]\psi^*(x)\psi(x-y)\\
&=\int\!\!\D x\,\psi^*(x)\left(\frac{1}{\I}\frac{\D}{\D x}\right)^n\psi(x)\,.
\end{aligned}
\label{ingold:eq:kmoment}
\end{equation}
A generalization to moments containing both position and momentum
operators is possible if a Weyl ordering for the operators is
respected (for further details see \cite{ingold:hille84}).

To start, it is instructive to consider a few special quantum
states of simple structure which also play a key role in
the discussion of disordered systems. Let us begin with a plane wave,
i.e. $\psi(x)=\exp(\I k_0 x)/\sqrt{2\pi}$. Inserting this wave
function into (\ref{ingold:eq:wigner}) one finds
\begin{equation}
W(x,k) = \int\frac{\D y}{2\pi}\E^{\I(k-k_0)y} = \delta(k-k_0)\,.
\label{ingold:eq:wpw}
\end{equation}
The intermediate result emphasizes the fact that the off-diagonal
contributions in (\ref{ingold:eq:wigner}), which are parameterized
by the coordinate $y$, contain the information about the momentum.
For a localized state, $\psi(x)=\delta(x-x_0)$, the Wigner
function emerges as
\begin{equation}
W(x,k) = \E^{2\I k(x-x_0)}\delta(x-x_0) = \delta(x-x_0)\,.
\label{ingold:eq:wloc}
\end{equation}
Next we consider a quantum state localized at two positions, i.e.
$\psi(x) =[\delta(x+a)+\delta(x-a)]/\sqrt{2}$. Proceeding as
before, one obtains the Wigner function
\begin{equation}
W(x,k) = \frac{1}{2}\left[\delta(x+a)+\delta(x-a)\right]
         +\cos(2ka)\delta(x)\,.
\label{ingold:eq:wtwoloc}
\end{equation}
The first two terms describe the localization of the particle at
$x=\pm a$. In addition, there occurs  an oscillatory term in the
middle between the two localization centers which accounts for the
coherent superposition of two localized states. This simple
example demonstrates that the Wigner function generally is not
positive even though it could be treated as a phase space density
in (\ref{ingold:eq:xmoment}) and (\ref{ingold:eq:kmoment}). This
characteristic feature is a direct consequence of the fact that 
by virtue of the definition (\ref{ingold:eq:wigner}) one attempts to 
define a phase space representation which allows for perfect localization 
in position and momentum as indicated by the results (\ref{ingold:eq:wpw}) 
and (\ref{ingold:eq:wloc}). This formal attempt to circumvent the
Heisenberg uncertainty relation is generally paid for by negative
parts of the Wigner function.

Thus far, we have considered a particle on a continuous and
infinitely extended one-dimensional state space. The situation
changes when we try to apply phase space concepts to the 
Anderson model of disordered systems. For such a lattice model the
integral in (\ref{ingold:eq:wigner}) has to be replaced by a sum.
Moreover, in numerical calculations, the system size has to be
taken finite. This results in a phase space that is twisted to a
torus so that additionally periodic boundary conditions appear in
momentum space apart from those usually imposed in real space.
Therefore, interference terms in momentum space analogous to those
appearing in real space as in (\ref{ingold:eq:wtwoloc}) occur also
across the boundaries of the Brillouin zone. Such artifacts can
also be understood as arising from the discretization of the
integral in (\ref{ingold:eq:wigner}) due to the lattice structure
of the model. In particular for the states located at the
boundaries of the Brillouin zone, the wave number $k$ is so large
that a discretized version of the Fourier integral no longer
represents a good approximation. The usage of Wigner functions for
lattice models therefore becomes problematic.

\begin{figure}[t]
\begin{center}
\subfigure{\includegraphics[width=\textwidth]{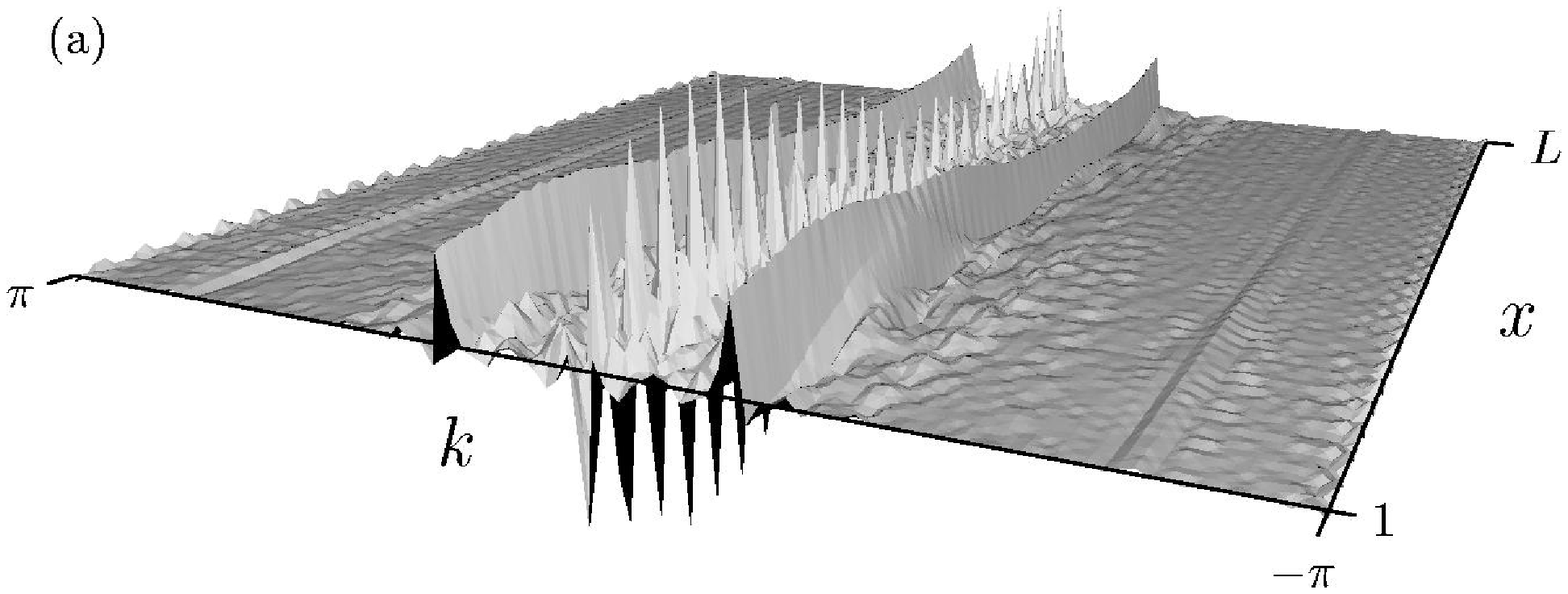}%
\label{ingold:fig:wigner}}

\subfigure{\includegraphics[width=\textwidth]{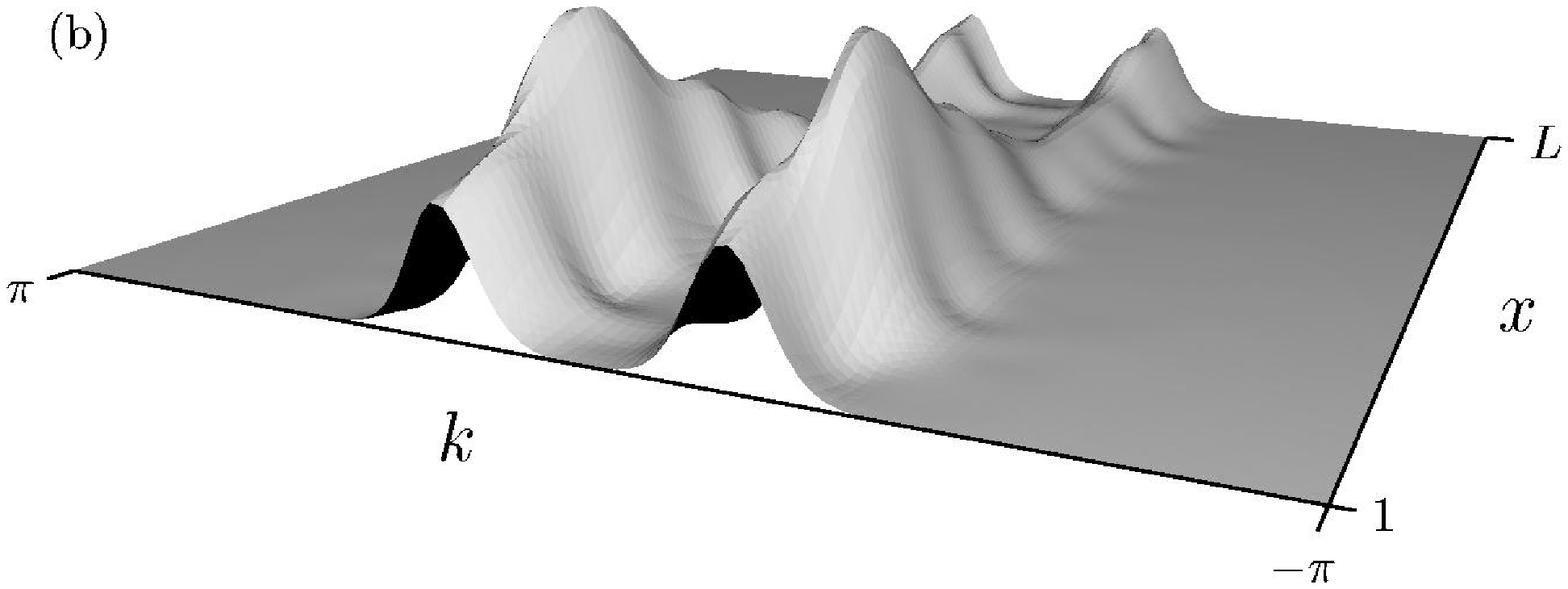}%
\label{ingold:fig:husimi}}
\end{center}
\caption{\textbf{(a)} Wigner function and \textbf{(b)} Husimi function for 
an eigenstate of the Anderson model}
\end{figure}

In Fig.~\ref{ingold:fig:wigner} we present an example for the
Wigner function of an eigenstate of the Anderson model at
relatively low disorder. While a certain spatial localization has
already set in, two dominant wave numbers $\pm k$ can still be
clearly recognized. At $k=0$, interference effects analogous to
those found in (\ref{ingold:eq:wtwoloc}) are visible while the two
low ridges at larger wave numbers represent the artifacts due to
the lattice structure of the Anderson model discussed in the
previous paragraph.

\subsection{The Husimi function}
The above discussed shortcomings of the Wigner function can be
circumvented by  use of the so-called Husimi function. The latter
one is obtained from the Wigner function by a Gaussian smearing
according to
\begin{equation}
\rho(x,k) = \frac{1}{\pi}\int\!\!\D x'\D k'
              \exp\left[-\frac{(x-x')^2}{2\sigma^2}-2\sigma^2(k-k')^2\right]
          W(x',k')\,.
\label{ingold:eq:rhow}
\end{equation}
Even though one might assume that one discards information by transforming 
the Wigner function $W(x,k)$ into the Husimi function $\rho(x,k)$, this is 
actually not the case as one can show by  inverting (\ref{ingold:eq:rhow}): 
The Wigner function can then be recast as
\begin{align}
\label{ingold:eq:wrho}
W(x,k) &=\\
&\int\frac{\D u\D v}{2\pi}\frac{\D x'\D k'}{2\pi}
         \exp\left[\frac{\sigma^2u^2}{2}+\frac{v^2}{8\sigma^2}
               +\I(x'-x)u+\I(k'-k)v\right]\rho(x',k')\,.\nonumber
\end{align}
The existence of the integrals is guaranteed by the inherent asymptotic decay 
of the Husimi function provided that the integrals over $x'$ and
$k'$ are evaluated first.

From a physical point of view, the most important aspect of the
Gaussian smearing introduced in (\ref{ingold:eq:rhow}) is that 
the Heisenberg uncertainty relation is now accounted for in a natural
way. The Husimi function will never localize a quantum state in
phase space beyond the limits set by the requirement $\Delta
x\Delta k\ge 1/2$. In return, the Husimi function yields a
nonnegative-valued phase space distribution. Indeed, by inserting
(\ref{ingold:eq:wigner}) into (\ref{ingold:eq:rhow}) one can
express the Husimi function in terms of the wave function
$\psi(x)$ as the squared quantity
\begin{equation}
\rho(x,k) = \left\vert\int\D x' \psi(x')\frac{1}{(2\pi\sigma^2)^{1/4}}
              \exp\left[-\frac{(x'-x)^2}{4\sigma^2}
                    -\I kx'\right]\right\vert^2\,,
\label{ingold:eq:husimi}
\end{equation}
which manifestly is nonnegative. According to
(\ref{ingold:eq:husimi}), the Husimi function may be viewed as the
projection of the wave function onto a minimal uncertainty state
or, using the language of quantum optics, onto a coherent state.
At this point, the width $\sigma$ is still a parameter which can
be chosen freely. We will fix its value later on when applying
these phase space concepts to the Anderson model.

Yet another aspect of the Gaussian smearing pertains to the
artifacts mentioned in the previous section for the Wigner
function on a lattice. While in (\ref{ingold:eq:wigner}) all
values of $y$ contribute, this is no longer the case for the
Husimi function where the convolution with a Gaussian restricts
the difference in the arguments of the two wave functions. In
fact, even the interference term appearing in the Wigner function
(\ref{ingold:eq:wtwoloc}) for a continuum model is strongly
suppressed in the corresponding Husimi function provided the distance 
$a$ is sufficiently larger than the width $\sigma$; this is evident from
\begin{equation}
\begin{aligned}
\rho(x,k) &= \frac{1}{2\sqrt{2\pi\sigma^2}}\bigg[%
                 \exp\left(-\frac{(x-a)^2}{2\sigma^2}\right)+
                 \exp\left(-\frac{(x+a)^2}{2\sigma^2}\right)\\
&\hspace{0.3\linewidth}+2\exp\left(-\frac{x^2}{2\sigma^2}-
                          \frac{a^2}{2\sigma^2}\right)\cos(2ka)\bigg]\,.
\end{aligned}
\end{equation}

A comparison of the Wigner function in
Fig.~\ref{ingold:fig:wigner} and the Husimi function in
Fig.~\ref{ingold:fig:husimi} demonstrates the effect of the
Gaussian smearing. The spatial localization and the two mainly
contributing wave numbers are the dominant features of the Husimi
function. The interferences leading to negative contributions to
the Wigner function have almost disappeared except for some
wiggles which, however, do not render the Husimi function negative. 
Finally, the features stemming from the lattice have now disappeared.

\subsection{Inverse participation ratio}
The Husimi function, in particular for quantum  states at
intermediate disorder strengths, does contain a rich structure.  In
general, it is desirable for various reasons to reduce this wealth
of information and to characterize the state by a single number
which is derived from the Husimi function. First, in clear
contrast to the one-dimensional case depicted in
Fig.~\ref{ingold:fig:husimi} it is generally not possible to
visualize the Husimi function for systems in two and higher space
dimensions. Second, the details of the Husimi functions depend
sensitively on the chosen disorder realization. In order to perform
averages, it is necessary to reduce the characterization of the
states to a number.

Since the Husimi function is nonnegative, in principle all
quantities familiar from classical phase space analysis can be
employed in the quantum case as well. One possibility is given by
the so-called Wehrl entropy \cite{ingold:wehrl79}
\begin{equation}
S=-\int\frac{\D x\D k}{2\pi}\rho(x,k)\ln[\rho(x,k)]\,,
\label{ingold:eq:wehrl}
\end{equation}
which has been used, e.g., in the discussion of the driven rotor
\cite{ingold:gorin97}. The above-mentioned existence of a mapping
between the kicked rotor and the Anderson model has motivated a
study of the latter by means of the Wehrl entropy
\cite{ingold:weinm99}.

From a numerical point of view, it is advantageous to linearize
the Wehrl entropy (\ref{ingold:eq:wehrl}). Replacing $\ln(x)$ by
its linear approximation $x-1$, one obtains $-x\ln(x)\approx
x-x^2$. The approximation shows qualitatively the some behavior
as the original function and, in particular, yields the same values 
at the boundaries at $x=0$ and $1$. Performing this linearization in
(\ref{ingold:eq:wehrl}), the first order term gives rise to a
constant due to normalization and we are left with the second
order term, the inverse participation ratio (IPR) in phase space
\begin{equation}
P = \int\frac{\D x\D k}{2\pi}[\rho(x,k)]^2\,.
\label{ingold:eq:ipr}
\end{equation}
While both the Wehrl entropy $S$ and the phase space IPR $P$ are
entirely determined in terms of the wave function $\psi(x)$, it is
necessary for the evaluation of the Wehrl entropy to first
calculate the Husimi function explicitly. This requires a
significant numerical effort which can be reduced by resorting to
the phase space IPR \cite{ingold:manfr00}. In turn, this makes
possible  the study of the Anderson model in two and even in three
dimensions \cite{ingold:wobst02}. Inserting
(\ref{ingold:eq:husimi}) into (\ref{ingold:eq:ipr}) one obtains
the phase space IPR expressed directly in terms of the wave function
as
\begin{equation}
P=\frac{1}{8\sqrt{\pi}\sigma}\int\!\D u\left\vert\int\!\D v\,
         \psi\left(\frac{u-v}{2}\right)\psi\left(\frac{u+v}{2}\right)
     \exp\left(-\frac{v^2}{8\sigma^2}\right)\right\vert^2\,.
\end{equation}
Apart from the numerical aspects, another advantage relates to the
fact that inverse participation ratios  may be defined as well in
real and momentum space, i.e.,
\begin{equation}
P_x=\int\!\D x\,\vert\psi(x)\vert^4\qquad\mathrm{and}\qquad
P_k=\int\!\D k\,\vert\widetilde\psi(k)\vert^4\,,
\end{equation}
where $\widetilde\psi(k)$ is the wave function in momentum
representation. The availability of the inverse participation
ratio in different spaces allows for instructive physical
comparisons. The real space IPR $P_x$ is a well studied quantity as
it is related to the return probability of a diffusing particle
\cite{ingold:thoul74} while the phase space IPR has recently been
used to describe the complexity of quantum states
\cite{ingold:sugit02,ingold:sugit01}. We add that a quantity containing
information similar to the phase space IPR can be defined on the
basis of marginal distributions in real and momentum space
\cite{ingold:varga02}.

In order to demonstrate that it is justified to refer to the
above-mentioned quantities as inverse participation ratios, we
consider a few special cases for quantum states on a one-dimensional 
lattice with $L$ lattice sites. Let us first assume that the state is 
localized on a single site. Then we find $P_x=1$ and $P_k=1/L$. 
The inverse of these results indeed indicates that in real space the 
state occupies only one site while in momentum space $L$ sites are 
occupied. In phase space, one obtains $P=1/2\sqrt{\pi}\sigma$ which 
primarily reflects the Gaussian width in real space.

As a second example, we consider a real-valued ballistic wave
function at momentum $\bar k$, i.e. $\psi(x)=\sqrt{2/L}\cos(\bar
kx)$. In real and momentum space, the inverse participation ratios
become $P_x=3/2L$ and $P_k=1/2$, respectively. The first result
accounts for the nonuniform extension in real space while the
latter indicates the contribution of the two momenta $\pm\bar k$.
For $\bar k\ne 0,\pm\pi$, the phase space IPR
$P=(\sigma\sqrt{\pi}/L)[1+2\exp(-4\sigma^2\bar k^2)]$ depends on
the distance between the two contributing momenta.

So far, we have not specified the width $\sigma$ entering both,
the definition (\ref{ingold:eq:husimi}) of the Husimi function and
the results for the phase space IPR $P$ mentioned in
the previous two paragraphs. Although one is rather free in
choosing $\sigma$, the most impartial choice consists in selecting
an equal relative resolution in real and momentum space, i.e.
$\Delta x/L=\Delta k/2\pi$. Together with the requirement of
minimal uncertainty, $\Delta x\Delta k=1/2$, one finds $\Delta
x=\sigma=\sqrt{L/4\pi}$. For our two examples mentioned above
(which will appear later in the limits of vanishing and very
strong disorder) this implies, that the IPR  $P$ will scale like
$1/\sqrt{L}$ as a function of system size. This result for
one-dimensional systems is found to generalize to $P\sim L^{-d/2}$
in $d$ dimensions.

\begin{figure}[t]
\begin{center}
\includegraphics[width=0.4\linewidth]{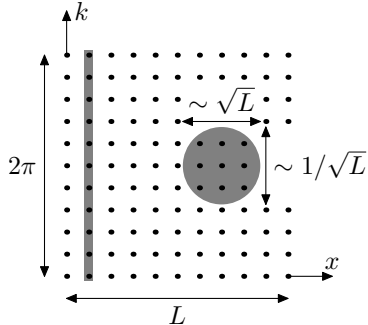}
\end{center}
\caption{The different phase space resolution provided by real space wave
function and Husimi function is visualized by the gray shaded areas on the
left and right, respectively}
\label{ingold:fig:uncert}
\end{figure}

An intuition for the differences between inverse participation
ratios in real and momentum space, on the one hand, and the IPR in
phase space, on the other hand, can be obtained by considering the
resolution provided by the wave function and the Husimi function
in real and momentum space. The vertical gray stripe on the left
in Fig.~\ref{ingold:fig:uncert} corresponds to the real space wave
function. Independent of system size, this wave function always
leads to perfect resolution in real space. The price to be paid,
however, consists in the absence of \textit{any} resolution in
momentum space. The converse holds true for the momentum space
wave function. The situation is quite different for the Husimi
function symbolized by the gray disk in
Fig.~\ref{ingold:fig:uncert}. Although the discussion can be
generalized \cite{ingold:ingol02}, we will consider the case where
$\sigma\sim\sqrt{L}$. Then, structures occurring on $\sqrt{L}$
lattice points, or less, cannot be resolved. While this resolution
becomes worse as the system size is increased, the resolution
relative to the system size improves with $1/\sqrt{L}$. For large
system sizes, the {\em phase space approach} therefore allows for
a detailed description of a quantum state both in real and
momentum space, while a specific wave function representation will
always  neglect one of the two possibilities.

\section{Anderson model in phase space}
In the preceding section we have demonstrated that a phase space approach
indeed presents advantages when both, real and momentum space
properties are of interest as it is the case for disordered
systems when the disorder strength is varied from zero to
infinity. Although, in principle, the Husimi function is
equivalent to the wave function, it will make the relevant
information more readily accessible. We next want to illustrate
this very fact by considering the Anderson model for disordered
systems. Its Hamiltonian \cite{ingold:ander58}
\begin{equation}
H = -t\sum_{<\mathbf{x},\mathbf{x'}>}(\vert\mathbf{x'}\rangle
\langle\mathbf{x}\vert+\vert\mathbf{x}\rangle\langle\mathbf{x'}\vert)
+W\sum_n v_n\vert\mathbf{x}\rangle\langle\mathbf{x}\vert
\label{ingold:eq:anderson}
\end{equation}
is defined on a $d$ dimensional square lattice with $L$ sites in
each direction. In order to avoid boundary effects, periodic
boundary conditions are imposed. The first term describes the
kinetic energy which allows for hopping between nearest neighbor
sites $<\mathbf{x},\mathbf{x'}>$. In the following, the hopping
matrix element will set the energy scale, i.e. $t=1$. The second
term on the right-hand side of (\ref{ingold:eq:anderson})
represents the disordered on-site potential where the energies
$v_n$ are independently drawn from a box distribution on
the interval $[-1/2;1/2]$. The respective disorder strength is then
determined by $W$.

\subsection{Husimi functions}

\begin{figure}[p]
\begin{center}
\includegraphics[width=\linewidth]{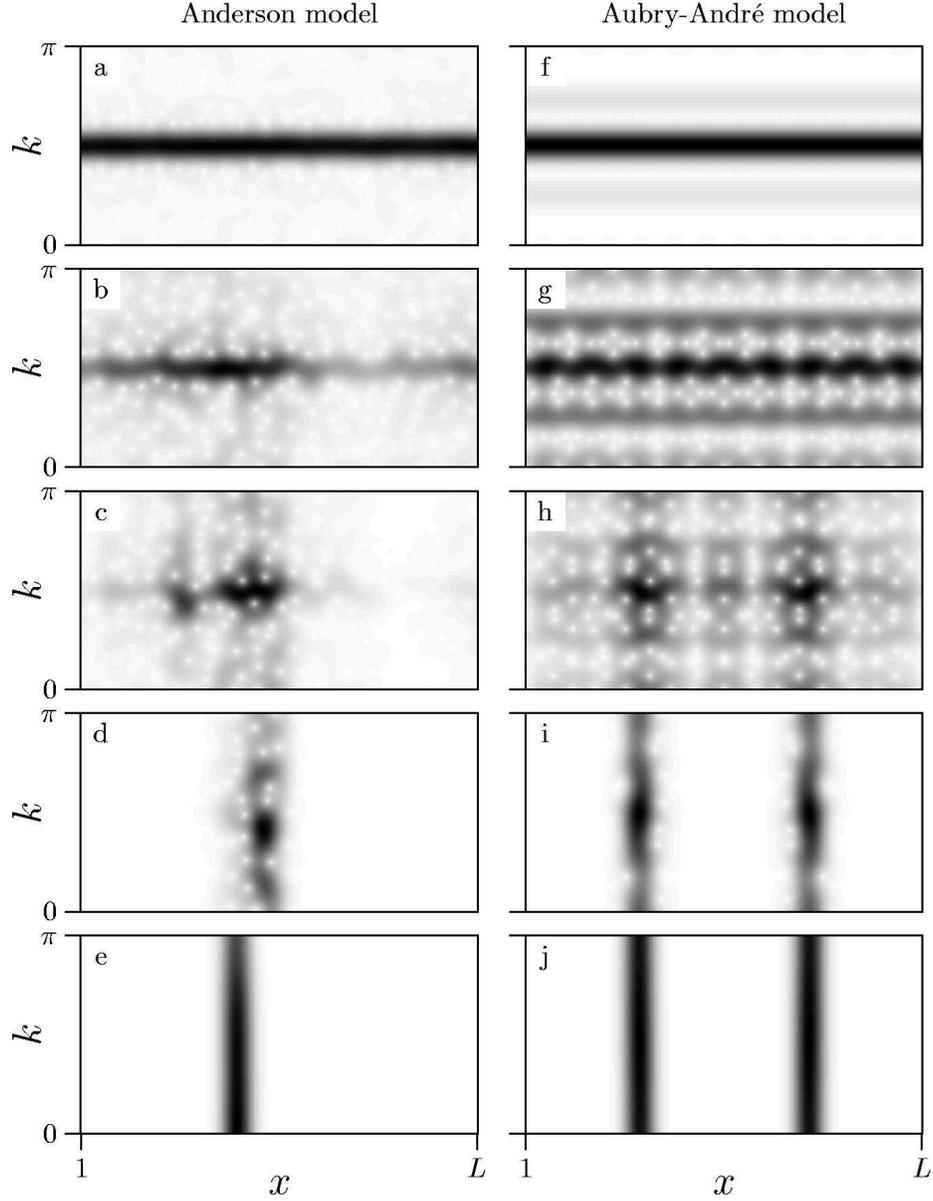}
\end{center}
\caption{Husimi functions for \textbf{(a--e)} the Anderson model with disorder
strength $W=0.1, 1, 2, 4$, and $40$ and \textbf{(f--j)} the Aubry-Andr{\'e} 
model with potential strength $\lambda=0.1, 1.5, 2, 2.5$, and  $10$ for a 
state close to the band center and a system size $L=377$. In view of the 
symmetry $\rho(x,-k)=\rho(x,k)$ only the upper half of the phase space is 
shown}
\label{ingold:fig:andaub}
\end{figure}

Before we investigate the inverse participation ratio for the
Anderson model, it is instructive to first take a look at the
underlying Husimi functions. Since a visualization is readily
possible only for one-dimensional models, we compare in
Fig.~\ref{ingold:fig:andaub} the one-dimensional Anderson model
and the Aubry-Andr{\'e} model \cite{ingold:aubry80}. The latter is
based on a periodic potential $\lambda\sum_n\cos(2\pi\beta n)\vert
n\rangle\langle n\vert$ replacing the disorder potential in
(\ref{ingold:eq:anderson}). Choosing $\beta$ as the golden mean,
$(\sqrt{5}-1)/2$, the potential is incommensurate with the
underlying lattice and displays a phase transition from
delocalized to localized states \cite{ingold:aubry80}. We have
chosen this model for comparison since its phase space IPR is
comparable to that obtained for the two- and three-dimensional
Anderson model \cite{ingold:ingol02}. In
Fig.~\ref{ingold:fig:andaub}, the Aubry-Andr{\'e} model therefore
serves as a substitute for the higher-dimensional Anderson models.

Before we discuss in further detail the Husimi functions depicted
in Fig.~\ref{ingold:fig:andaub}, we mention that for real wave
functions the Husimi function, according to its definition
(\ref{ingold:eq:husimi}), respects the symmetry $\rho(x,k) =
\rho(x,-k)$. It is therefore sufficient to plot the Husimi
functions solely for positive momenta from which the lower
part may be reconstructed as a mirror image with respect to $k=0$.

The Husimi functions for the limiting cases of very weak and very
strong potential depicted in Figs.~\ref{ingold:fig:andaub}(a) and 
\ref{ingold:fig:andaub}(e) for the Anderson model and 
Figs.~\ref{ingold:fig:andaub}(f) and \ref{ingold:fig:andaub}(j)
for the Aubry-Andr{\'e} model can readily be interpreted in terms of states
localized either in momentum or in real space. In addition, in
Fig.~\ref{ingold:fig:andaub}(f) the coupling between plane waves due
to the periodic potential becomes visible while the localization
at two sites seen in Fig.~\ref{ingold:fig:andaub}(j) arises because
the calculation was restricted to the class of antisymmetric
states \cite{ingold:thoul83}. Furthermore, we keep the state number
fixed while passing through avoided crossings so that states may be
localized at different positions as a function of potential strength (cf.\
e.g. Figs.~\ref{ingold:fig:andaub}(d) and \ref{ingold:fig:andaub}(e)).

More interesting than the interpretation of the Husimi function in 
the limiting cases is a discussion of the transition from plane waves 
for a weak potential to spatially localized states for a strong potential. 
The one-dimensional Anderson model and the Aubry-Andr{\'e} model display 
two different scenarios with the latter being also characteristic for
higher-dimensional Anderson models (cf.\ Sect.~\ref{ingold:subsec:andipr}).

For the one-dimensional Anderson model, the Husimi function for a
plane wave (Fig.~\ref{ingold:fig:andaub}(a)) contracts in real space
(Fig.~\ref{ingold:fig:andaub}(b)) and displays a very well localized
core at intermediate disorder strength $W$
(Fig.~\ref{ingold:fig:andaub}(c)). A further increase of the
potential strength leads to a spreading in momentum direction
(Fig.~\ref{ingold:fig:andaub}(d)) and finally a state localized in real 
space (Fig.~\ref{ingold:fig:andaub}(e)) is approached.

For the Aubry-Andr{\'e} model the scenario is quite different. As
mentioned above, already a weak potential strength $\lambda$ can lead 
to a coupling between plane waves of very different momenta 
(Fig.~\ref{ingold:fig:andaub}(f)). This type of coupling is accompanied by an 
increased filling of phase space
(Fig.~\ref{ingold:fig:andaub}(g)) up to  potential strengths
$\lambda$ located just below the critical value of $\lambda=2$
where the localization transition takes place
(Fig.~\ref{ingold:fig:andaub}(h)). There, for increasing system size,
a more and more abrupt contraction in phase space occurs, cf.\ 
Fig.~\ref{ingold:fig:andaub}(i). Finally, for very strong potentials, 
the Husimi function of a localized antisymmetric state depicted in 
Fig.~\ref{ingold:fig:andaub}(j) forms.

\subsection{Inverse participation ratios}
\label{ingold:subsec:andipr}
While the Husimi functions depicted in
Fig.~\ref{ingold:fig:andaub} nicely elucidate  the phase space
behavior of a state as a function of potential strength, a more
quantitative analysis is desirable. On the basis of inverse participation
ratios, a study of the higher-dimensional Anderson model becomes feasible
and averages over disorder realizations can be performed. This in turn
will allow us to draw general conclusions about the physics that rules the 
localization transition in the Anderson model.

\begin{figure}[t]
\begin{center}
\includegraphics[width=0.9\textwidth]{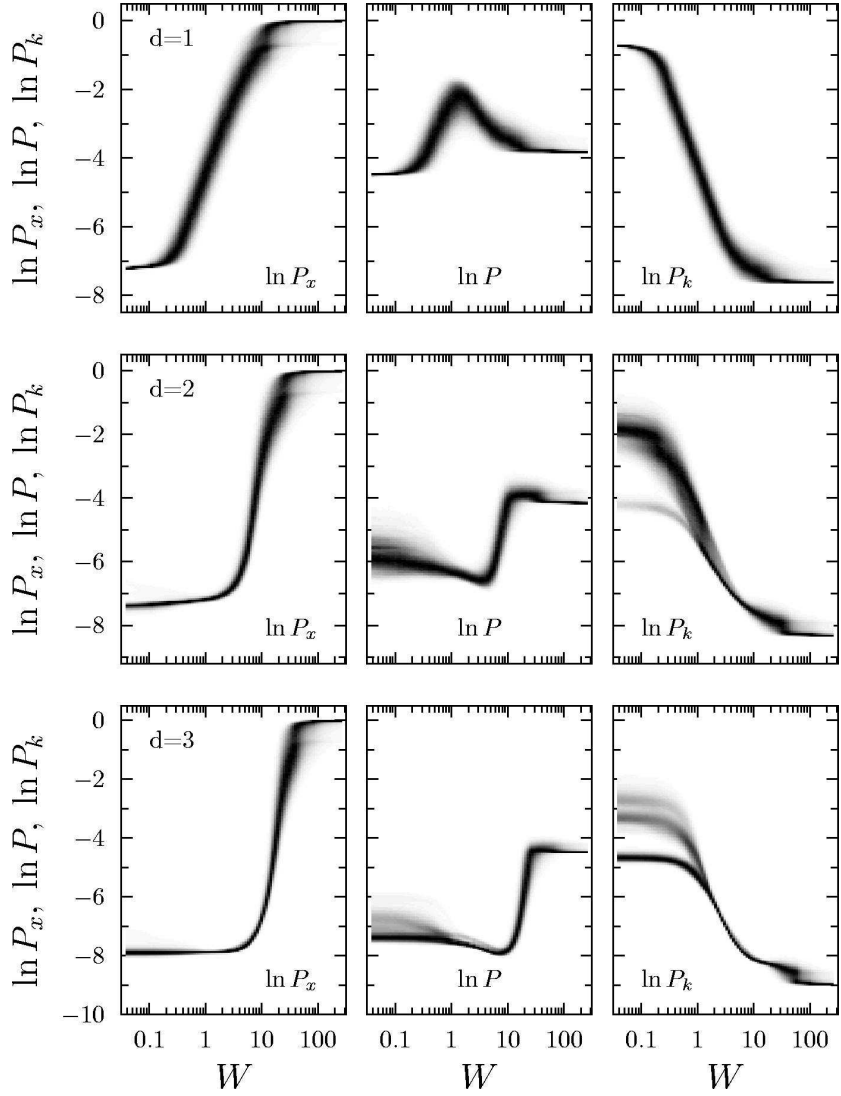}
\end{center}
\caption{Distribution of inverse participation ratios in real space (left),
phase space (middle), and momentum space (right) for the Anderson model in
$d=1$ (upper row, $L=2048$), $d=2$ (middle row, $L=64$), and $d=3$ (lower
row, $L=20$)}
\label{ingold:fig:ipr}
\end{figure}

In Fig.~\ref{ingold:fig:ipr} we present distributions of the inverse
participation ratios $P_x$, $P$, and $P_k$ in real space, phase space, and
momentum space, respectively, for the Anderson model in one, two, and three
dimensions. The distributions have been obtained by diagonalizing the
Anderson model for 50 disorder realizations in $d=1$ and $2$ and taking
half of the eigenstates around the bandcenter for each realization. In $d=3$
it was sufficient to consider only 20 disorder realizations.

Upon comparing the real space IPRs for the three different
dimensions, no qualitative differences can be seen. The overall
behavior shows an increase of $P_x$ (left column) with increasing 
potential strength reflecting the spatial localization of the wave
functions. Correspondingly, the momentum space IPR (right column) 
decreases with increasing potential strength.

In clear contrast, the phase space IPR behaves qualitatively different
for the Anderson model in $d=1$ and $d\ge2$. For $d=1$, the peak in 
$P$ is consistent with our observations for the Husimi function in
Figs.~\ref{ingold:fig:andaub}(a--e), namely the contraction of the
Husimi function at intermediate disorder strength.

For $d=2$ and $3$, the behavior of the phase space IPR rather
corresponds to what we have observed for the Husimi function of
the Aubry-Andr{\'e} model in Figs.~\ref{ingold:fig:andaub}(f--j).
With increasing potential strength, the Husimi function spreads in
phase space implying a decrease of $P$. Then, at the
transition, a sudden contraction occurs, implying a jump to
relatively large values of $P$. This scenario can be understood in 
physical terms as follows. The large spread in phase space is associated
with the presence of a diffusive regime which is known to exist in $d=2$ 
and larger. This interpretation is corroborated by a comparison with 
results from energy level statistics \cite{ingold:wobst02}. The jump in
$P$, on the other hand, indicates the Anderson transition
\cite{ingold:abrah79,ingold:lee85}. Here, one might
object that the case $d=2$ is the marginal case where strictly
speaking no phase transition occurs. One finds indeed, that both, the
minimum and the maximum of $P$, shift towards vanishing disorder as the
system size is increased \cite{ingold:wobstxx}. In contrast, in $d=3$, the 
minimum of $P$ shifts to larger disorder strength $W$, while the maximum 
shifts in the opposite direction, in agreement with the existence of the
Anderson transition. 

A comparison of the Husimi functions for the one-dimensional
Anderson model and the Aubry-Andr{\'e} model reveals the
differences between the Anderson model in $d=1$ and in higher
dimensions. In the one-dimensional Anderson model, a weak disorder
potential predominantly couples a plane wave to other,
energetically almost degenerate plane waves. Therefore, the 
coupling occurs between plane waves of almost the same momentum.
Due to the finite resolution in phase space, this coupling cannot
be observed in momentum space. In real space, however, the
coupling results in large scale variations of the Husimi function
which in the end cause a contraction in phase space as discussed
above. In clear contrast, for the Aubry-Andr{\'e} model the situation 
is quite different: Here,  the coupling predominantly occurs to distant 
momentum values as can be seen in Fig.~\ref{ingold:fig:andaub}(f). 
This finally leads to a spreading of the Husimi function in phase space 
\cite{ingold:ingol02}. The very same scenario is valid for the Anderson 
model in two and higher dimensions. Again, there exist energetically
almost degenerate momenta which are very different from those present
in the original plane wave. The coupling due to a weak disorder potential
thus again leads to a spreading in phase space.
On the other hand, perturbation theory for strong disorder shows that the 
limiting value of the phase space IPR for large $W$ is approached from 
above \cite{ingold:wobstxx}. Therefore, a jump from small to large values 
of $P$ arises and thus a phase transition (in $d=3$) is to be expected 
at an intermediate disorder strength $W$.

In conclusion, these considerations in phase space in terms of concepts 
such as the Husimi function and the corresponding inverse participation 
ratio prove indeed very valuable in order to explore in greater detail 
the physics for a class of quantum systems which is so dear to Bernhard 
Kramer, namely disordered quantum systems \cite{ingold:krame93}.

\section*{Acknowledgment}
The authors have enjoyed constructive and insightful discussions
with S.~Kohler, I.~Varga, and D.~Weinmann. This work was supported
by the Sonderforschungsbereich 484 of the Deutsche
Forschungsgemeinschaft. The numerical calculations were partly
carried out at the Leibniz-Rechenzentrum M{\"u}nchen. Moreover, two of
us (P.H., G-L.I.) are eagerly looking forward to see many more
insightful and provoking  achievements by Bernhard Kramer; he is
still young and vivacious enough to contribute to great science.

%

\end{document}